%% file: ExpParamCuts_v22.tex
\documentclass[preprint, 5p]{elsarticle}
 
\newcommand{\theolab}[1]{\label{theo:#1}}                   
\newcommand{\lemlab}[1]{\label{lem:#1}}                     
\newcommand{\corolab}[1]{\label{coro:#1}}                     
\newcommand{\seclab}[1]{\label{sec:#1}}                     
\newcommand{\eqlab}[1]{\label{eq:#1}}                         
\newcommand{\figlab}[1]{\label{fig:#1}}                         
\newcommand{\EQ}[1]{(\ref{eq:#1})}               
\newcommand{\FIG}[1]{Figure~\ref{fig:#1}}       
\newcommand{\LEMMA}[1]{Lemma~\ref{lem:#1}}       
\newcommand{\THEO}[1]{Theorem~\ref{theo:#1}}     
\newcommand{\CORO}[1]{Corollary~\ref{coro:#1}}     
\newtheorem{theo}{Theorem}[section]        
\newtheorem{lemma}[theo]{Lemma}            
\newtheorem{coro}[theo]{Corollary}            

\def\qed{\hskip 6pt \openbox ht{7pt}wd{4.3pt}}

\def\openbox ht#1wd#2{\hbox{\vrule\vbox{\hrule\vskip #1 \hbox{\null\hskip #2}
    \hrule}\vrule}}
\def\Proof:{\noindent{\sl{\bf Proof:\enspace}}}
\def\Proofof:#1{\noindent{\sl{\bf Proof (of #1):\enspace}}}

\def\nscr{{\cal N}}

\def\lam{\lambda}

\def\phi{\phi}
\def\newtheta{\theta}
\usepackage{amssymb,amsmath}
\long\def\comment#1{}
\usepackage{graphicx}
\usepackage[dvipsnames]{xcolor}
  \usepackage{tikz}
  
  \newcommand{\smallRightIndent}{\hspace{-2mm}&}
  \newcommand{\R}{\mathbb R}
 \usepackage[textsize=tiny,textwidth=1.0cm]{todonotes}
\presetkeys{todonotes}{inline}{}
\definecolor{darkRed}{RGB}{214,0,0}
\definecolor{darkGreen}{RGB}{5,121,24}
\definecolor{darkBlue}{RGB}{2,39,198}
\newcommand\red[1]{\textcolor{darkRed}{#1}}
\newcommand\blue[1]{\textcolor{darkBlue}{#1}}
\newcommand\green[1]{\textcolor{darkGreen}{#1}}

\newcommand{\vcom}[1]{}
\newcommand{\mcom}[1]{}
\newcommand{\tcom}[1]{}
\renewcommand{\phi}{\gamma}
  
\begin{document}

\begin{abstract} 
	There are many applications of max flow with capacities that depend
on one or more parameters.  Many of these applications fall into the
``Source-Sink Monotone" framework, a special case of Topkis's
monotonic optimization framework, which implies that the parametric
min cuts are nested.  When there is a single parameter, this property implies
that the number of distinct min cuts is linear in the number of
nodes, which is quite useful for constructing algorithms to identify all possible min cuts.

When there are multiple Source-Sink Monotone parameters, and vectors of parameters are ordered in the usual vector sense, the resulting min cuts are still
nested.  However, the number of distinct min cuts was an open question.  We show that even with only two parameters, the number of distinct min cuts can be exponential in
the number of nodes.  
\end{abstract}

\begin{keyword}
Network flow, Max Flow/Min Cut, Parametric flow
\end{keyword}

\begin{frontmatter}
	\title{Complexity of Source-Sink Monotone 2-Parameter Min Cut}
	
	\author[maxwell]{Maxwell Allman}
	\ead{mallman@stanford.edu}
	
	\author[venus]{Venus Lo}
	\ead{venus.hl.lo@cityu.edu.hk}
	
	\author[tom]{S. Thomas McCormick}
	\ead{tom.mccormick@sauder.ubc.ca}

	\address[maxwell]{Management Science and Engineering, Stanford University, Stanford, CA 94305, United States}
	
	\address[venus]{Department of Management Sciences, City University of Hong Kong, Kowloon, Hong Kong SAR}
	
	\address[tom]{Sauder School of Business, University of British Columbia, Vancouver, BC V6T 1Z2, Canada}
	
\end{frontmatter}

\section{Introduction} \seclab{intro}

There are many problems that can be formulated as Max Flow/Min Cut networks such that the capacities of the arcs depend on one or more {\em parameters}.  This problem is known as {\em parametric Max Flow/Min Cut}, see e.g.\cite{paramext} for a recent survey.  In such applications, we may be interested in identifying the parameters which maximize the flow value, or which lead to the zero of an auxiliary function.

Typically, the capacities of the arcs are affine functions of the
parameters (but see \cite{GusfieldMartel} for some exceptions).  In
this setting, it is easy to see that the region of the parameter
space where a specific min $s$--$t$ cut (hereafter min cut) is
optimal, which we call a {\em cell}, is a polyhedron.  The
interiors of these cells are disjoint and so induce a partition of the parameter space. If
the number of cells for some class of instances is ``small'' (polynomial in the size of the network), then an algorithm could try to construct all of the cells and efficiently solve the parametric problems in this class.  However, if the number of cells can be exponential in the size of the networks,
this algorithmic strategy is ruled out.  This makes it interesting to
study the {\em complexity} of classes of parametric networks, namely
the worst-case number of cells in the class.  In general parametric networks, Carstensen \cite{carst} showed that the number of cells can be exponential in the number of nodes in the network, even with just one parameter. 

On the other hand, Gallo, Grigoriadis, and Tarjan \cite {ggt} (hereafter GGT) showed that a class of parametric networks with many applications has only a {\em linear} number of cells.  Suppose that there is a single scalar parameter $\lam$, and that all capacities are affine in $\lam$.
We write an arc with tail $i$ and head $j$ as $i\to j$, and its (upper) capacity as $u_{ij}(\lam) = a_{ij}\lam + c_{ij}$ ($\lam$ is implicitly restricted to a domain where all capacities are
non-negative).  The nodes of the network are the {\em source} $s$, the {\em sink} $t$, and the {\em internal nodes} $N$ with $n = |N|$. Following \cite{paramext}, we say that a parametric network is {\em Source-Sink Monotone (SSM)} if
  \begin{align} 
  u_{sj}(\lam) &\text{ is non-decreasing in $\lam$ for all arcs exiting $s$} \nonumber \\ &\text{(i.e., $a_{sj}\ge 0$);} \nonumber \\
  u_{it}(\lam) &\text{ is non-increasing in $\lam$ for all arcs entering $t$}\nonumber \\ &\text{(i.e., $a_{it}\le 0$);} \nonumber \\
  u_{ij}(\lam) &\text{ is constant in $\lam$ for all other arcs}\nonumber \\ &\text{(i.e., for all $i\ne s$ and $j\ne t$, $a_{ij} = 0$).}        \eqlab{SSM}
  \end{align}
GGT showed that SSM networks have at most $n$ cells, and so at most $n$ distinct min cuts.  This is quite convenient for construction of algorithms, and indeed another major contribution of GGT is an algorithm that computes all
of the min cuts in the same asymptotic time as a single max flow computation.  These results were generalized to larger classes of networks with a single parameter in \cite{paramext}, and to parametric
min-cost flow networks in \cite{MCFparam}.

There are applications where multiple parameters arise naturally.  One broad class is problems with multiple objectives (see \cite{ehrgott} for a survey), where it is natural to model $k$ objectives by introducing $k-1$ parameters to obtain another model with a single objective equal to a linear combination of the original $k$ objectives.  Other examples include a scheduling application of Chen \cite{chen} and a budgeted network interdiction problem \cite{budgetnet}.  Thus it is interesting to extend the GGT-type results to multiple parameters.

Notice that the definition of SSM extends naturally to multiple
parameters if we replace ``$\lam$'' with ``$\lam_1, \dots, \lam_k$''.
In fact, the GGT result which proves that single-parameter SSM
networks have at most $n$ cells is a special case of a more general
result on monotone parametric optimization on lattices by Topkis
\cite{topkis,TopBook}.  When specialized to SSM parametric networks
with parameters $\lam_1$, $\lam_2$, \dots, $\lam_k$, Topkis's
framework looks like this: Let $S(\lam_1, \dots, \lam_k)$ denote a min
cut at $(\lam_1, \dots, \lam_k)$.  If
$(\lam_1, \dots, \lam_k) \le (\lam'_1, \dots, \lam'_k)$ (in the usual
partial order on real vectors), then
$S(\lam_1, \dots, \lam_k) \subseteq S(\lam'_1, \dots, \lam'_k)$ and
the min cuts are {\em nested}.  This result gives a quick proof for
why single-parameter SSM networks have at most $n$ min cuts: The
partial order on real vectors is a {\em total} order on $\R^1$, which
implies that all of the min cuts are nested and there can be at most
$n$ min cuts.

When we extend from one SSM parameter to multiple SSM parameters, we
still have nested min cuts due to Topkis's result.  However, we only
have a partial order on vectors of parameters instead of a total
order.  What is the complexity of parametric max flow/min cut
in this case?

The problem of the complexity of parametric {\em global} min cut with
$k$ parameters was studied in \cite{multcritmincut,Kar16}, who showed
that there are only $O(n^{k+1})$ cells, which is polynomial for fixed
$k$.   Given that global min cut and min $s$--$t$ cut seem to be
closely related, it is perhaps surprising that our main result is:

\begin{theo}  \theolab{main}
  There exists instances of SSM max flow/ min cut with two parameters where all $2^n$ $s$--$t$ cuts are unique min cuts for some  values of the parameters.
\end{theo}

 We will prove \THEO{main} by constructing a family of instances,
 parameter values, and flows that show that every possible $s$--$t$ cut is a
 unique min cut at some parameter values.  For simplicity we will henceforth denote our two parameters as $\lam$ and $\mu$, and so the capacity of $i\to j$ is $a_{ij}\lam + b_{ij}\mu + c_{ij}$.  We further specialize the SSM conditions \EQ{SSM} to:
\begin{align} 
  u_{sj}(\lam, \mu) &\text{ is increasing in $\lam$ and $\mu$ for all arcs exiting $s$} \nonumber
   \\ & \text{(i.e., $a_{sj}, b_{sj} > 0$);} \nonumber \\
  u_{ij}(\lam, \mu) &\text{ is constant in $\lam$ and $\mu$ for all other arcs} \nonumber 
   \\ & \text{(i.e., for all $i\ne s$, $a_{ij} = b_{ij} = 0$).}  \eqlab{sSSM}
\end{align}
We consider only non-negative values for $\lam$ and $\mu$.  In our construction, the constant term $c_{sj}$ in the parametric arcs with tail $s$ is always zero, and hence the capacities of all arcs $s\to j$ are non-negative for all $\lam, \mu\ge 0$.

In Section \ref{sec:construct},  we present the recursive definitions for the networks, parameters values, and flows.  In Section \ref{sec:intuition}, we motivate these definitions and give the intuition behind the construction.   Section \ref{sec:proofs} proves a series of lemmas which lead to the proof of Theorem \ref{theo:main}.  Finally, Section \ref{sec:discuss} discusses some implications of our result.
      
\section{Construction of the Family of Examples} \seclab{construct}

For each $n\ge 1$ we construct a parametric network $\nscr^n$
satisfying our restricted notion of SSM in \EQ{sSSM}.  Let
$N^n = \{1, \dots, n\}$ denote the internal nodes of $\nscr^n$.  The
set of arcs of $\nscr^n$ is the union of {\em source arcs} $\{s \to j:\; j \in N^n\}$, {\em sink arcs}
$\{j \to t:\; j \in N^n\}$, and {\em internal arcs} $\{j \to k: \; j, k \in N^n, j > k\}$.  The capacity of arc $s \to j$ is $u^n_{sj}(\lam, \mu) = a^n_{sj}\lam + b^n_{sj}\mu\comment{+
  c^n_{sj}}$.  The constructed values of $a^n_{sj}$ and $b^n_{sj}$ are
positive, which ensures that $\nscr^n$ is SSM.  The capacity of a
non-parametric arc $j \to k$ is $u^n_{jk}$ for $j \neq s$.

\vcom{Removed $S\backslash \{i\} = S-i$ since it's never used.}
For $S \subset N^n$ such that $i\notin S$, we denote $S\cup \{i\}$ by
$S+i$.  Any $S\subseteq N^n$ induces the $s$--$t$ cut $S + s$, but for
simplicity we refer to $S$ as the cut.  For each $S\subseteq N^n$, we
construct a point $(\lam^{n, S},\mu^{n,S})\in (0,1)\times (0,1)$ and a
corresponding flow $x^{n,S}$ such that $x^{n,S}$ is a feasible max
flow in $\nscr^n$ with unique min cut $S$ at
$(\lam^{n,S}, \mu^{n,S})$.  To reduce clutter, let $u_{sj}^{n,S}$
denote the capacity of $s \to j$ at $(\lam^{n, S},\mu^{n,S})$:
\begin{align*} 
u_{sj}^{n,S}=u^n_{sj}(\lam^{n,S}, \mu^{n,S}) = a^n_{sj} \lam^{n,S} +
b^n_{sj}\mu^{n,S}.
\end{align*}
The capacities of the other arcs are constant.  

\comment{Our construction will lead to some arcs $s\to j$ with $c^n_{sj} < 0$,
which would imply a negative capacity when $\lam = \mu = 0$.  This is
easy to fix using a standard trick: Define
$\alpha^n_j = (c^n_{sj})^+$, and add $\alpha^n_j$ to both $c^n_{sj}$
and $u^n_{jt}$.  This increases the capacity of every cut by
$\sum_j \alpha^n_j$ and so min cuts do not change, and it makes all
$c^n_{sj}$ non-negative.}

Our construction is recursive so that $\nscr^n$ is based on $\nscr^{n-1}$.
For $S\subseteq N^{n-1}$, we construct $(\lam^{n,S},\mu^{n,S})$ and
$(\lam^{n,S+n},\mu^{n,S+n})$ based on $(\lam^{n-1,S}, \mu^{n-1,S})$,
and we construct $x^{n,S}$ and $x^{n,S+n}$ based on $x^{n-1,S}$.

For the base case at $n = 1$, let $a^1_{s1} = b^1_{s1} = 1$\comment{, $c^1_{s1} = 0$} so that the arc capacities are:
\begin{align*}
u^1_{s1}(\lam, \mu) &= \lam + \mu, \text{ and } u^1_{1t} = 1.
\end{align*}
For $S = \emptyset$, set:
\begin{align*}
(\lam^{1,\emptyset}, \mu^{1,\emptyset}) = \left(\frac{1}{4}, \frac{1}{4} \right), \text{ with }
x^{1,\emptyset}_{s1} = \frac{1}{2} = x^{1,\emptyset}_{1t}.
\end{align*}  
Clearly, the flow $x^{1,\emptyset}$ is feasible and $S = \emptyset$ is the unique min cut.  For $S = \{1\}$, set:
\begin{align*}
(\lam^{1,\{1\}}, \mu^{1,\{1\}}) = \left(\frac{3}{4}, \frac{3}{4} \right), \text{ with } x^{1,\{1\}}_{s1} = 1 = x^{1,\{1\}}_{1t}.
\end{align*} 
Clearly, the flow $x^{1,\{1\}}$ is feasible and $S = \{1\}$ is the unique min cut.  Hence, $S=\emptyset, \{1\}$ are unique min cuts at
$(\lam^{1,\emptyset}, \mu^{1,\emptyset})$ and
$(\lam^{1,\{1\}}, \mu^{1,\{1\}})$ respectively.

We define two ``large'' constants for $n\ge 2$ whose
definitions will be motivated later:  
\begin{align}
     \newtheta^n & =  3a^{n-1}_{s,n-1},  \text{ and} \eqlab{thdef} \\
  \phi^{n} & = \begin{cases} 4 & \text{for $n = 2$}, \\
      4\cdot\sum_{j < n} (\newtheta^n b^{n-1}_{sj} -
       3a^{n-1}_{sj}) & \text{for $n>2$.} 
     \end{cases}  \eqlab{phidef} 
\end{align}
To construct $\nscr^n$ from $\nscr^{n-1}$ for $n>1$, we add node $n$,
arcs $s\to n$ and $n\to t$, and arcs $n \to j$ for $j \in N^{n-1}$.  The arc capacities for $\nscr^n$ are as follows:
\begin{align}
u^n_{sj}(\lam,\mu) & =  4 a^{n-1}_{sj} \lam + (1 + \newtheta^n) b^{n-1}_{sj} \mu \smallRightIndent \text{ for } &j\in N^{n-1}; \nonumber\\
u^n_{jt} & =  u^{n-1}_{jt} + \newtheta^n b^{n-1}_{sj} \smallRightIndent \text{ for }& j\in N^{n-1};  \nonumber\\
u^n_{ij} & =  u^{n-1}_{ij} \smallRightIndent  \text{ for }&  i, j\in N^{n-1}, i > j; \nonumber \\
u^n_{nj} & =  \newtheta^n b^{n-1}_{sj} -3a^{n-1}_{sj} \smallRightIndent  \text{ for }& j\in N^{n-1}; \nonumber \\
u^n_{sn}(\lam, \mu) & =  \phi^{n} \lam + \mu; &\nonumber  \\
u^n_{nt} & =  \frac{\phi^{n}}{2}. &\eqlab{capacities}
\end{align}
By \EQ{thdef}, we have $u_{n, n-1}^n=0$.  Furthermore, we have
$a_{sj}^n = 4 a_{sj}^{n-1}$ and $b_{sj}^n = (1+\newtheta^n)
b_{sj}^{n-1}$ for $j \in N^{n-1}$, and $a^n_{sn} =\phi^n$ and $b^n_{sn} = 1$.  We will verify that $u^n_{nj} \ge 0$ for $j <n$ in \LEMMA{phi}.  See Figures \ref{fig:2cells} and \ref{fig:3cells} for $\nscr^2$ and $\nscr^3$.

\begin{figure}
	\begin{center}
		\scalebox{.43}{\includegraphics{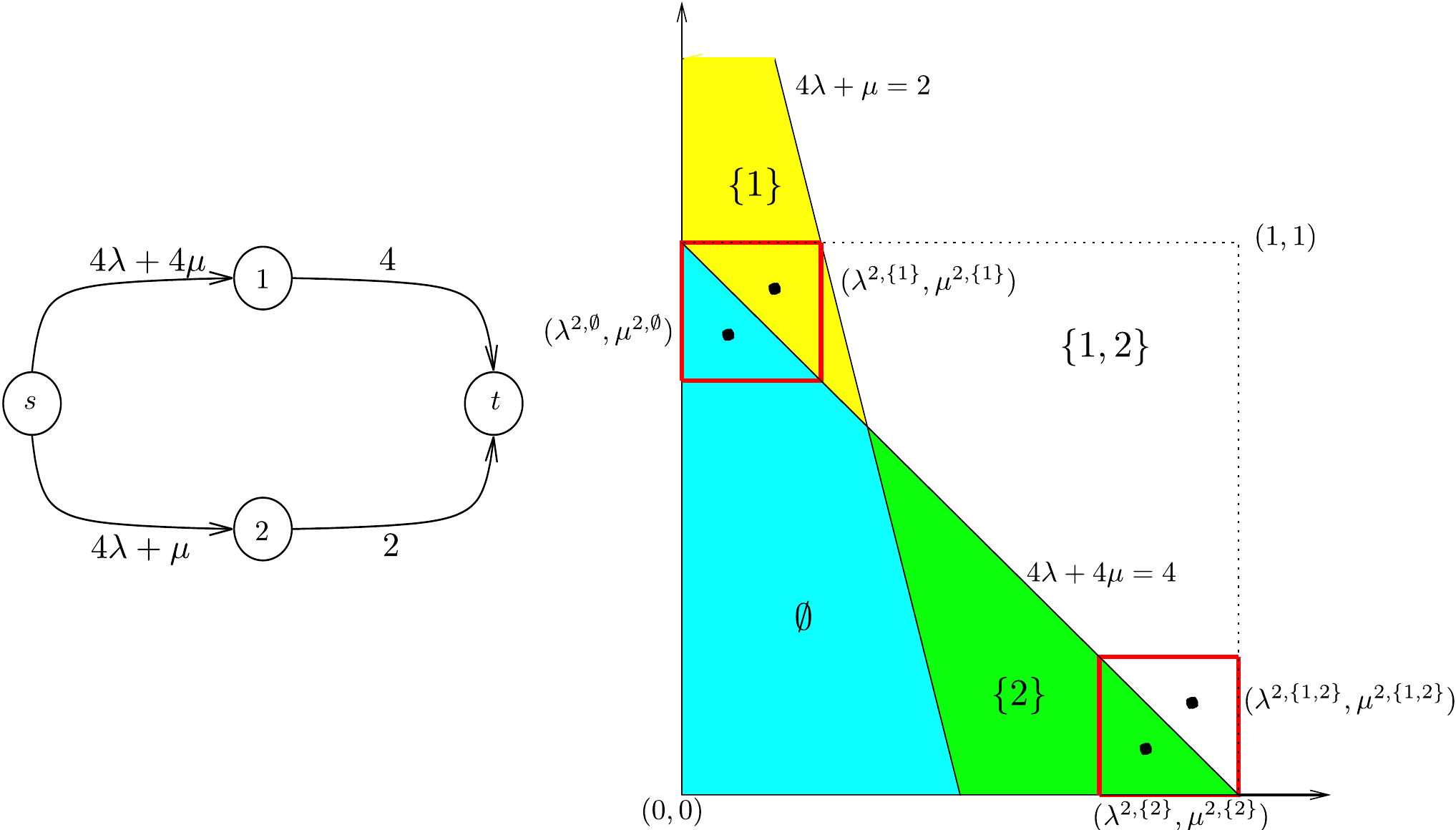}} 
		\caption{For
			$n=2$, $\newtheta^2 = 3$ and $\phi^{2} = 4$, which leads to this
			parametric network $\nscr^2$ and its four cells.  The two
			scaled, shifted boxes (outlined in red) are
			$(0,\frac{1}{4})\times (\frac{3}{4}, 1)$, and
			$(\frac{3}{4},1)\times (0,\frac{1}{4})$.  The four
			$(\lam^{2,S},\mu^{2,S})$ points are marked, and the dotted box
			$(0,1)\times (0,1)$ contains the two red boxes
			and the four points. We omit arcs with capacity 0 in the network.}
		\figlab{2cells}
	\end{center}
\end{figure}
\begin{figure}[hbt]
	\begin{center}
		\scalebox{.35}{\includegraphics{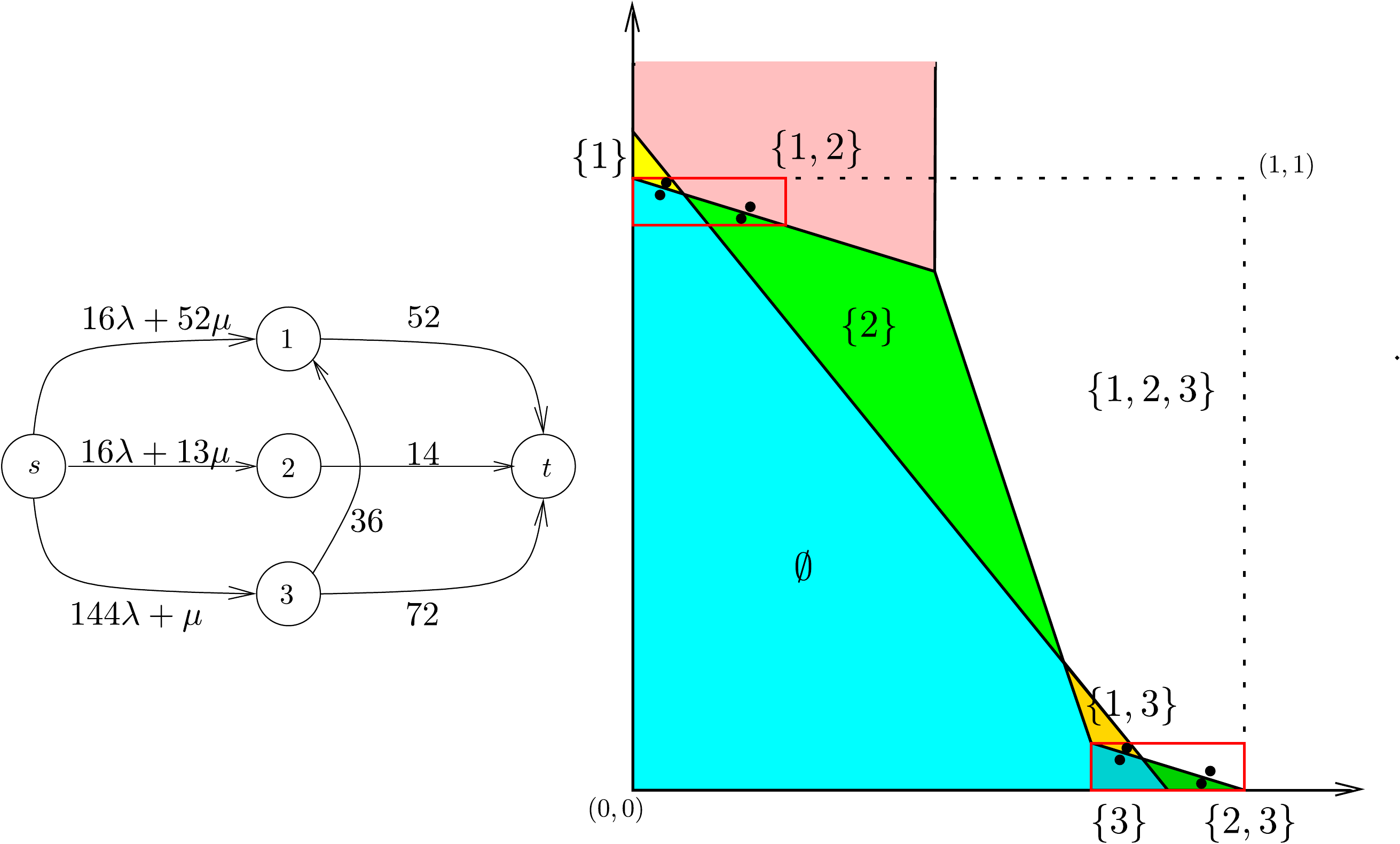}}
		\caption{For $n=3$, $\newtheta^3 = 12$ and $\phi^{3} = 144$, which leads to this
			parametric network $\nscr^3$ and its eight cells.  The two
			scaled, shifted boxes (outlined in red) are
			$(0,\frac{1}{4})\times (\frac{12}{13}, 1)$, and
			$(\frac{3}{4},1)\times (0,\frac{1}{13})$, which contain the
			eight $(\lam^{3,S},\mu^{3,S})$ points (marked but not
			labeled).  Each red box contains a scaled,
			shifted copy of the cells for $n=2$.  The dotted box
			$(0,1)\times (0,1)$ contains the two red boxes and the eight
			points.  We omit arcs with capacity 0 in the network.} 
		\figlab{3cells}
	\end{center}
\end{figure}

Let $S \subseteq N^{n-1}$.  First, we define the following point at
which $S$ will be the unique min cut of the new $\nscr^n$:
\begin{align}
(\lam^{n,S}, \mu^{n,S}) & = \left(\frac{\lam^{n-1,S}}{4},
\frac{\mu^{n-1,S} + \newtheta^n}{1+\newtheta^n}\right) .
 \tag{6a} \eqlab{newdata_without}
\end{align}
Since each point $(\lam^{n-1,S}, \mu^{n-1,S})\in (0,1)\times (0,1)$,
it is easy to see that 
$(\lam^{n,S}, \mu^{n,S})\in (0, \frac{1}{4})\times
(\frac{\newtheta^n}{1+\newtheta^n}, 1)$.  The corresponding flow $x^{n,S}$ is:  
\begin{align}
x^{n,S}_{sj} & =  x^{n-1,S}_{sj} + \newtheta^n b^{n-1}_{sj} & \text{ for }& j\in N^{n-1};\nonumber \\
x^{n,S}_{jt} & =  x^{n-1,S}_{jt} + \newtheta^n b^{n-1}_{sj} & \text{ for }& j\in N^{n-1}; \nonumber\\
x^{n,S}_{ij} & =  x^{n-1,S}_{ij} & \text{ for }&  i, j\in N^{n-1}, i > j; \nonumber\\
x^{n,S}_{nj} & =  0 & \text{ for }& j\in N^{n-1}; \nonumber \\
x^{n,S}_{sn} &= x^{n,S}_{nt}  =  u^{n,S}_{sn} .&
\tag{6b} \eqlab{newflow_without}
\end{align}
Lemma \ref{lem:xnon} will use this flow $x^{n,S}$ to prove that 
$S$ is the unique min cut at $(\lam^{n,S}, \mu^{n,S})$.

Next, we define the following point at which $S+n$ will be the unique min cut of $\nscr^n$:
\begin{align}
(\lam^{n,S+n}, \mu^{n,S+n}) & = \left(\frac{\lam^{n-1,S} + 3}{4}, \frac{\mu^{n-1,S}}{1+\newtheta^n}\right).
\tag{7a}
\eqlab{newdata_with}
\end{align}
Observe that 
$(\lam^{n,S+n}, \mu^{n,S+n})\in (\frac{3}{4},1)\times (0,
\frac{1}{1+\newtheta^n})$.  The corresponding flow $x^{n,S+n}$ is: 
\begin{align}
x^{n,S+n}_{sj} & =  x^{n-1,S}_{sj} +  3a^{n-1}_{sj} \smallRightIndent \text{ for } &j\in N^{n-1};  \nonumber \\
x^{n,S+n}_{jt} & =  x^{n-1,S}_{jt} + \newtheta^n b^{n-1}_{sj} \smallRightIndent \text{ for } &j\in N^{n-1};  \nonumber \\
x^{n,S+n}_{ij} & =  x^{n-1,S}_{ij}  \smallRightIndent \text{ for }& i, j\in  N^{n-1}, i > j;  \nonumber \\
x^{n,S+n}_{nj} & =  u^n_{nj} = \newtheta^n b^{n-1}_{sj} - 3a^{n-1}_{sj} \smallRightIndent \text{ for }& j\in N^{n-1};  \nonumber \\
x^{n,S+n}_{sn} & =  \frac{\phi^{n}}{2} + \sum_{j<n} (\newtheta^n b^{n-1}_{sj} - 3a^{n-1}_{sj});&  \nonumber \\
x^{n,S+n}_{nt} & =  \frac{\phi^{n}}{2}. 
\tag{7b} \eqlab{newflow_with}
\end{align}
Lemma \ref{lem:xyesn} will use this flow $x^{n,S+n}$ to prove that 
$S+n$ is the unique min cut at $(\lam^{n,S+n}, \mu^{n,S+n})$.

%

\section{Intuition behind the Construction} \seclab{intuition}

Strong complementary slackness implies that $S$ is the unique min cut
at $(\lam^{n,S}, \mu^{n,S})$ if there exists a flow $x^{n,S}$ (which
then must be a max flow) satisfying: 
\begin{align*} 
x^{n,S}_{sj} & <  u^{n,S}_{sj} & \text{ for }& j\in S;  \\
\blue{x^{n,S}_{jt}} & \blue{{}=  u^{n}_{jt}} & \blue{\text{ for }} &\blue{j\in S};   \\
\blue{x^{n,S}_{sj}} & \blue{{}=  u^{n,S}_{sj}} & \blue{\text{ for }} &\blue{j\notin S};    \tag{SCS}  \\ 
x^{n,S}_{jt} & <  u^{n}_{jt}  & \text{ for } &j\notin S; \\  
\green{x^{n,S}_{jk}} & \green{{}=  u^{n}_{jk}} & \green{\text{ for }} &\green{j\in S, k\notin S,  j > k};   \\
\red{x^{n,S}_{jk}} & \red{{}=  0} & \red{\text{ for }} &\red{j\notin S, k\in S,  j> k}. 
\end{align*} 
See \FIG{nscr_minus} for an example and explanation of the colors.
Lemmas \ref{lem:xnon} and \ref{lem:xyesn} will use (SCS) to prove
Theorem \ref{theo:main}. 
\comment{By giving the construction of $\nscr^n$, $x^{n,S}$, and
$(\lam^{n,S}, \mu^{n,S})$, and proving that $x^{n,S}$ and $S$ satisfy
(SCS) at $(\lam^{n,S}, \mu^{n,S})$ for every $S \subseteq N^n$, we
establish each of the $2^n$ possible $s$--$t$ cuts in $\nscr^n$ as the
unique min cut for some values of $\lam$ and $\mu$.  Hence, the
complexity of 2-parameter SSM max flow/min cut is exponential.  }

\begin{figure}
	\begin{center}
		\scalebox{.5}{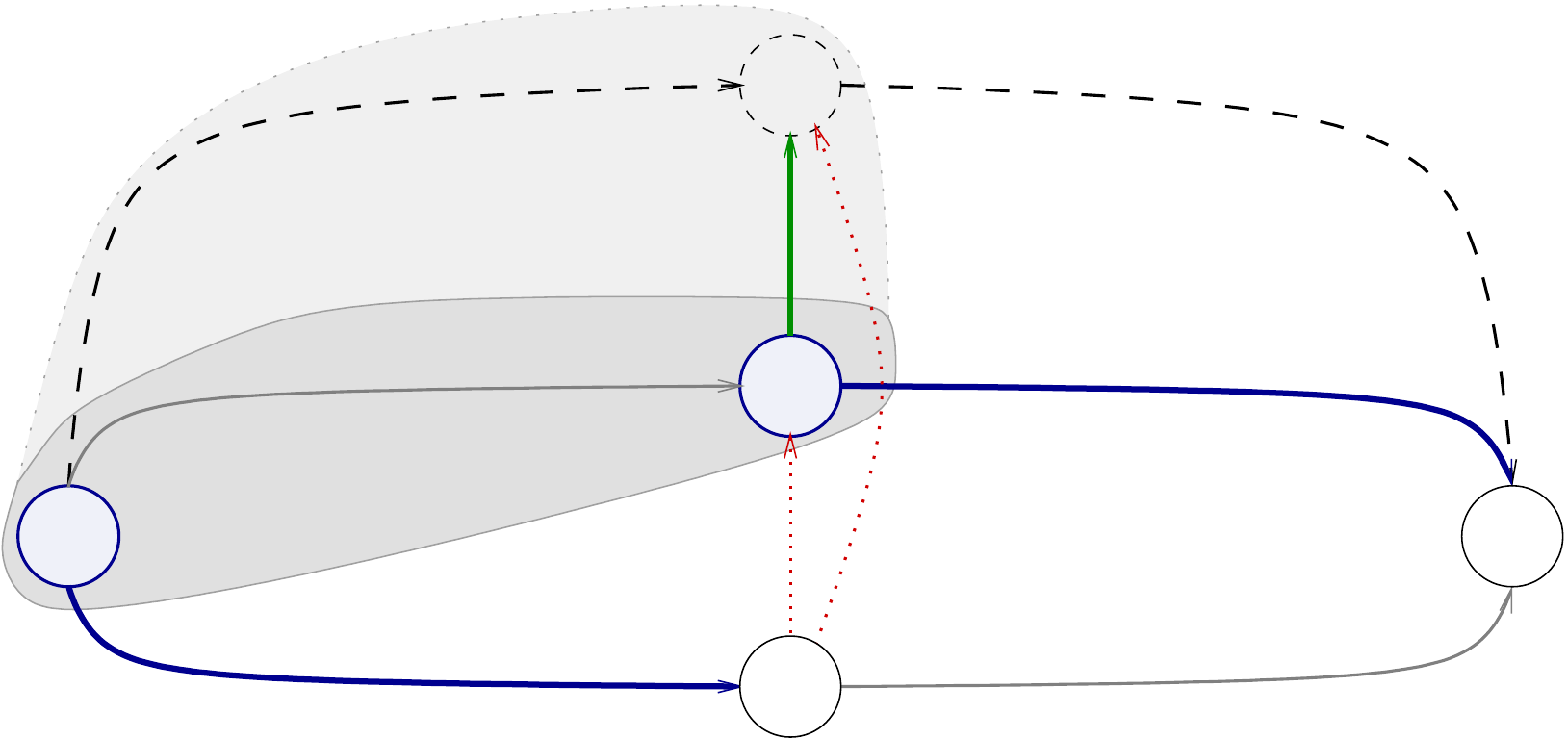}
		\caption{$\nscr^{n-1}$, denoting $x^{n-1}_{ij}$ by
			$x^-$ and $u^{n-1}_{ij}$ by $u^-$.  The (SCS)
			conditions say that if $i \notin S$, then $s \to i$
			is saturated (heavy \blue{blue}), but $i \to t$ is
			not saturated, and all other $i \to k$ have 0 flow
			(dotted \red{red}); whereas if $j \in S$, then
			$s\to j$ is not saturated, but $j \to t$ is
			saturated, and all other $j\to k$ are saturated
			(heavy \green{green}).  The dashed source and sink
			arcs at $k$ indicate that they are saturated
			depending on whether or not $k$ is in $S$; the (SCS)
			conditions hold for internal arcs with head $k$ in
			either case.}
		\figlab{nscr_minus}	
	\end{center}
\end{figure}

First consider the base case of $\nscr^1$.  It is easy to check that our flows $x^{1, \emptyset}$ and $x^{1, \{1\}}$ satisfy (SCS).  If we visualize the
cuts as cells on the $(\lambda, \mu)$-plane, then we have a box on
$(0,1) \times (0,1)$ which contains both points
$(\lam^{1,\emptyset}, \mu^{1,\emptyset})=(\frac{1}{4}, \frac{1}{4})$ and
$(\lam^{1,\{1\}}, \mu^{1,\{1\}})=(\frac{3}{4}, \frac{3}{4})$.

For $n>1$, our strategy is to shrink the original box $(0,1)\times (0,1)$ containing all the $(\lam^{n-1,S}, \mu^{n-1,S})$ points into the smaller box $(0, \frac{1}{4})\times (\frac{\newtheta^n}{1+\newtheta^n}, 1)$.  Then for each $S \subseteq N^{n-1}$, the points corresponding to min cuts without node $n$, $(\lam^{n,S}, \mu^{n,S})$, are contained in the box $(0, \frac{1}{4})\times (\frac{\newtheta^n}{1+\newtheta^n}, 1)$.  We also construct a new point for each min cut which contains node $n$, $(\lam^{n,S+n}, \mu^{n,S+n})$, such that the new points are contained in the box $(\frac{3}{4},1)\times (0, \frac{1}{1+\newtheta^n})$.  The boxes containing the min cuts with $n$ and without $n$ do not overlap.  

\setcounter{equation}{7}

When we add node $n$ to $\nscr^n$, the arc capacities of $s \to n$ and $n \to t$ are $u^n_{sn}(\lambda, \mu) =  \phi^{n} \lambda + \mu$ and $u^n_{nt} = \frac{\phi^{n}}{2}$.  When $\lam < \frac{1}{4}$, then $u^{n}_{sn}(\lambda, \mu)$ is ``small''
relative to $u^n_{nt}$ and so any max flow will saturate arc $s\to n$.  Hence, node $n$ will not be in any min cut and we can leave the flow on
internal arcs $n\to j$ as 0.  For each $S \subseteq N^{n-1}$ and $j <n$, \EQ{capacities} and \EQ{newdata_without} imply that:
\begin{align}  
u^{n,S}_{sj} & =   4a^{n-1}_{sj}\left(\frac{\lam^{n-1,S}}{4}\right) + (1 +\newtheta^n)b^{n-1}_{sj} \left(\frac{\mu^{n-1,S} +\newtheta^n}{1+\newtheta^n}\right) \nonumber\\
& =  a^{n-1}_{sj} \lam^{n-1,S} + b^{n-1}_{sj} \mu^{n-1,S} + \newtheta^n b^{n-1}_{sj} \nonumber \\
& =  u^{n-1,S}_{sj} + \newtheta^n b^{n-1}_{sj}.  \eqlab{Snon} 
\end{align}
\comment{The last equality above is true by the definition of $u^{n-1,S}_{sj}$.}  
Our construction adds $\theta^n b_{sj}^{n-1}$ to $u^{n-1,S}_{sj}$ and $u^{n-1}_{jt}$ to obtain $u^{n,S}_{sj}$ and $u^n_{jt}$.  By increasing the flow on $s \to j$ and $j \to t$ by the same amount of $\theta^n b_{sj}^{n-1}$, and having no flow on $n \to j$, we effectively create a copy of each cut $S \subseteq N^{n-1}$ (see Figure \ref{fig:nscr_notInS}).  We shrink and shift the
original box $(0,1)\times (0,1)$ into the top left corner.  Thus for each
$S\subseteq N^{n-1}$, the upper left corner of $(0,1)\times (0,1)$
(namely $(0, \frac{1}{4})\times (\frac{\newtheta^n}{1+\newtheta^n}, 1)$)
contains a cell for every min cut $S$.  See Figure \ref{fig:2cells} and upper left corner of Figure \ref{fig:3cells}.

\begin{figure}
	\begin{center}
		\scalebox{.5}{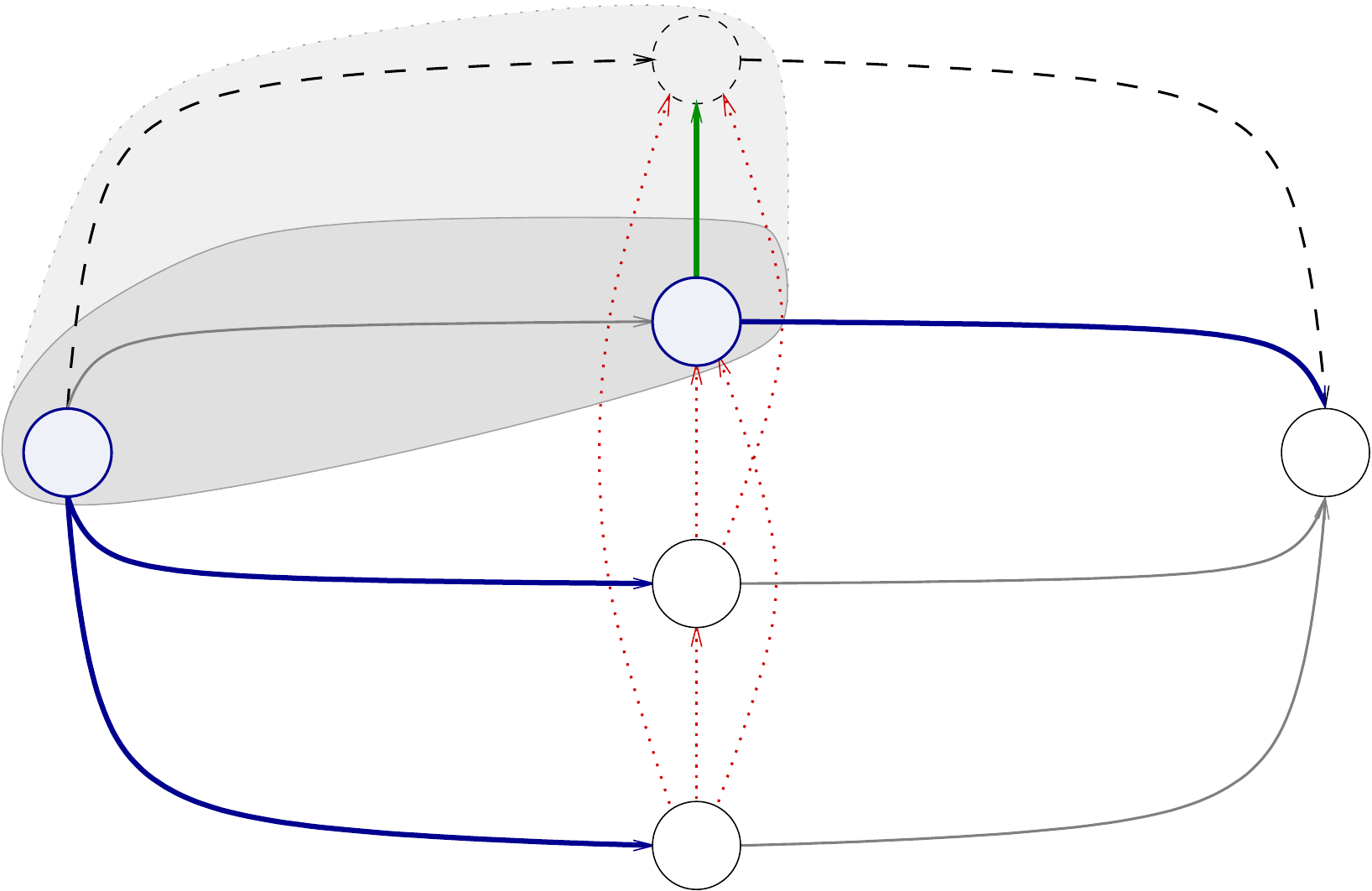}
		\caption{$\nscr^{n}$ with $n \notin S$; see
			\FIG{nscr_minus} for corresponding $\nscr^{n-1}$,
			explanation of node $k$, and key to the colors.  We
			denote $x^{n-1}_{ij}$ by $x^-$, $u^{n-1}_{ij}$ by
			$u^-$, $b^{n-1}_{sj}$ by $b^-$, $x^n_{ij}$ by $x^+$,
			$u^n_{ij}$ by $u^+$, $\newtheta^n$ by $\newtheta$,
			and $\phi^{n}$ by $\phi$.  When we add node $n$ and
			$\lam < \frac{1}{4}$, $s \to n$ is saturated and
			$n \to t$ is not saturated.  Arcs $n \to j$ have 0
			flow for all $j \in N^{n-1}$.  This maintains the
			saturated/non-saturated status on the other arcs.
			Hence, the previous min cut remains a min cut.  }
		\figlab{nscr_notInS}
	\end{center}
\end{figure}

On the other hand, when
$\lam >\frac{3}{4}$, then $u^{n}_{sn}(\lambda, \mu)$ is ``big'' relative to
$u^n_{nt}$ and so any max flow will saturate arc $n\to t$.  Hence, $n$ will
belong to every min cut and $\phi$ is big enough to ensure that we 
saturate internal arcs $n\to j$.  For each $S \subseteq N^{n-1}$ and $j <n$, \EQ{capacities} and \EQ{newdata_with} imply that:
\begin{align}
u^{n,S+n}_{sj} & = 4a^{n-1}_{sj}\left(\frac{\lam^{n-1,S}+ 3}{4}\right) + (1 + \newtheta^n)b^{n-1}_{sj} \left(\frac{\mu^{n-1,S}}{1+\newtheta^n}\right) \nonumber \\
& =  a^{n-1}_{sj} \lam^{n-1,S} + b^{n-1}_{sj} \mu^{n-1,S}  + 3a^{n-1}_{sj} \nonumber \\
& =  u^{n-1,S}_{sj} + 3a^{n-1}_{sj}.\eqlab{Syesn}
\end{align}
In this case, our construction adds $3a^{n-1}_{sj}$ to $u^{n-1,S}_{sj}$ to obtain $u_{sj}^{n, S+n}$, whereas it adds $\theta^n b_{sj}^{n-1}$ to $u^{n-1}_{jt}$ to obtain $u_{jt}^{n}$.  The saturating flow on $n\to j$ exactly makes up for the {\em update gap} between the (small) extra capacity added to $s\to j$ and the (big) extra capacity added to $j\to t$.  Saturated arcs in $\nscr^{n-1}$ remain saturated in $\nscr^n$, and unsaturated arcs in $\nscr^{n-1}$ remain unsaturated in $\nscr^n$.  We effectively create a copy and add $n$ to each cut $S \subseteq N^{n-1}$ (see Figure \ref{fig:nscr_InS}).  We shrink and shift the box containing $S+n$ into the bottom right corner.  Notice that $(\lam^{n,S+n}, \mu^{n,S+n}) = (\lam^{n,S}, \mu^{n,S}) +(\frac{3}{4}, -\frac{\newtheta^n}{1+\newtheta^n})$.  Thus for each $S\subseteq N^{n-1}$, the lower right corner of $(0,1) \times (0,1)$
(namely $(\frac{3}{4},1)\times (0, \frac{1}{1+\newtheta^n})$) contains a
cell for every min cut $S+n$.  See Figure \ref{fig:2cells} and
 lower right corner of Figure \ref{fig:3cells}.  This construction accounts for all $2^n$ min cuts of $\nscr^n$.

 \tcom{I took out this paragraph about integrality, ok?} 
 \comment{ By induction, observe that
 $\phi^{n}$ is always even, and hence $\newtheta^n$, $\phi^{n}$, $a^n_{sj}$,
 $b^n_{sj}$, and $u^n_{ij}$ (for $i\ne s$) are integers.  The
 values of $(\lam^{n,S}, \mu^{n,S})$ are always rational, but not necessarily integers.  Hence, the capacities of the arcs $u^{n,S}_{sj}$ and the
 flows $x^{n,S}_{ij}$ are often rational.}

\begin{figure}
	\begin{center}
		\scalebox{.5}{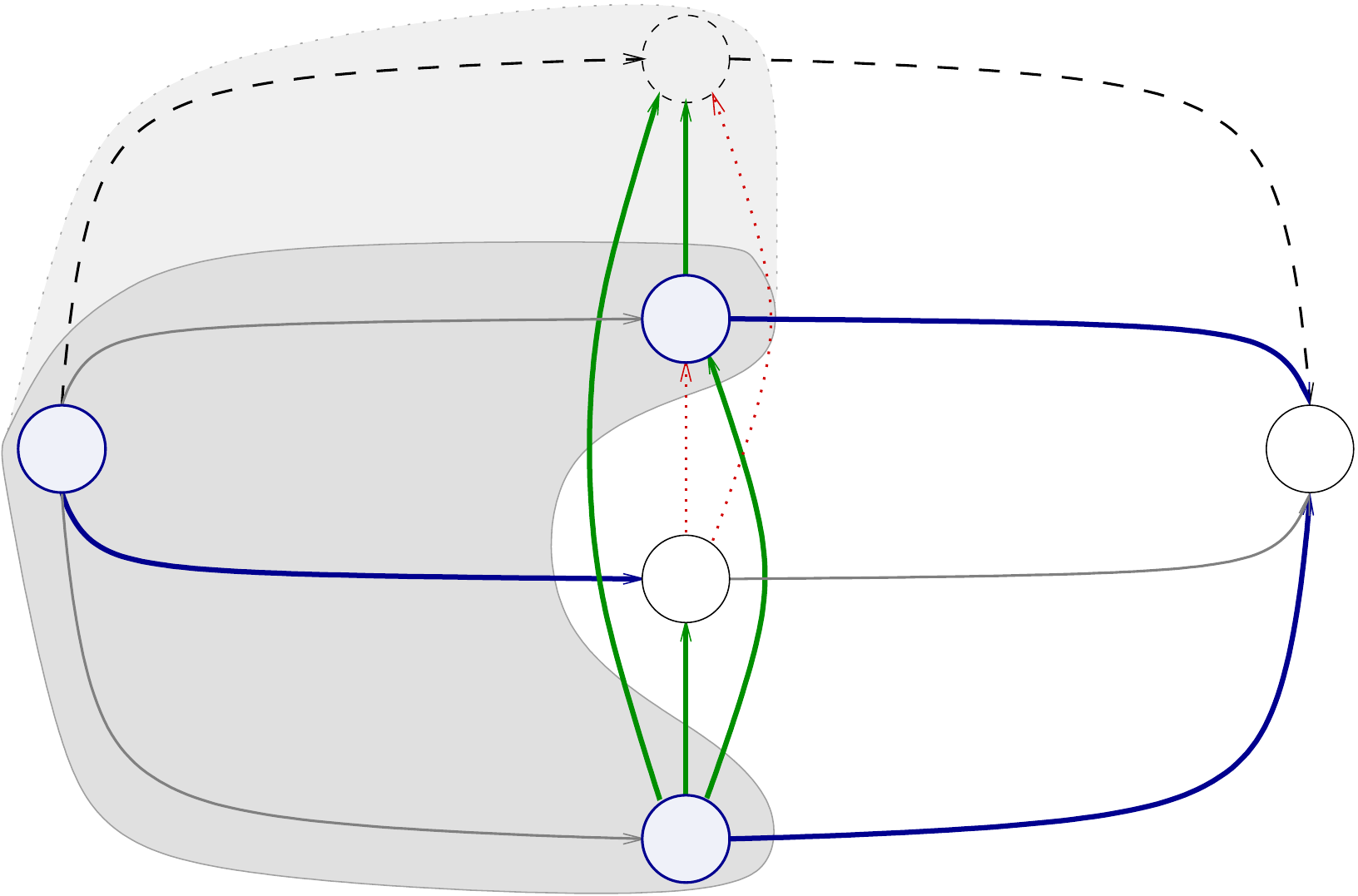}
		\caption{$\nscr^{n}$ with $n \in S$; see
			\FIG{nscr_minus} for corresponding $\nscr^{n-1}$,
			explanation of node $k$, and key to the colors, and
			\FIG{nscr_notInS} for notation.  When we add node
			$n$ and $\lam > \frac{3}{4}$, $s \to n$ is not
			saturated and $n \to t$ is saturated.  Arcs
			$n \to j$ are saturated with the amount of flow
			needed to make up the update gap between the
			added flow/capacity on $s\to j$, and the added
			flow/capacity on $j\to t$. This maintains the
			saturated/non-saturated status on the other arcs.
			Hence the new min cut is $S+n$.  }
		\figlab{nscr_InS}
	\end{center}
\end{figure}

\section{Proof of Correctness} \seclab{proofs}

Before proving that $x^{n,S}$ is a feasible flow at $(\lambda^{n,S}, \mu^{n, S})$ that satisfies (SCS), we prove the following technical lemma on $\phi^{n}$, which will be useful in subsequent proofs: 
\begin{lemma} \lemlab{phiGrowth}
	For $n \geq 3$, we have $\phi^{n} \geq  3(\phi^{n-1})^2$.
	\comment{Moreover, XXX $\phi^{n} = 4 \sum_{j<n} (\newtheta^n b^{n
			-1}_{sj} - 3 a^{n -1}_{sj})>4$.  } 
\end{lemma}

\Proof: We use induction on $n$.  For $n=3$, we
have $\phi^{3} = 144 \geq 3(4)^2 = 3 (\phi^{2})^2$ (see \FIG{3cells}).

For $n \ge 4$, we use the facts that $u^n_{n,n-1} = 0$;
$b^{n-1}_{sj} = (1 + \newtheta^{n-1}) b^{n-2}_{sj}$, which implies that
$\newtheta^{n-1}b^{n-2}_{sj} = b^{n-1}_{sj} - b^{n-2}_{sj}$;
$a^{n-1}_{sj} = 4a^{n-2}_{sj}$; and
$\newtheta^n = 3\phi^{n-1}$.  By induction, $\phi^{n-1}$ is strictly increasing and $\newtheta^n = 3a^{n-1}_{s,n-1} = 3\phi^{n-1}\ge 12 > 4$:
\begin{align}
3 (\phi^{n-1})^2 &= \newtheta^n\phi^{n-1}  =  4\sum_{j < n-1} \left(\newtheta^n\newtheta^{n-1}b^{n-2}_{sj}-	3\newtheta^na^{n-2}_{sj}\right) \nonumber \\
& = 4\sum_{j<n} (\newtheta^n b^{n-1}_{sj} - 3a^{n-1}_{sj}) \nonumber  \\ & \quad - 4\sum_{j<n-1}\left (\newtheta^nb^{n-2}_{sj} - 3a^{n-1}_{sj} +	3\newtheta^na^{n-2}_{sj}\right) \nonumber \\ 
& = \phi^{n} - 4\sum_{j<n-1} \big[ \newtheta^nb^{n-2}_{sj} +	3a^{n-2}_{sj}(\newtheta^n - 4)\big]   \eqlab{key} \\
& \le  \phi^{n}.  \qed \nonumber
\end{align}
\comment{We have $\phi^{n} \geq 3 (\phi^{n-1})^2 > 48$, which implies that
	$\phi^{n} = 4 \sum_{j<n} (\newtheta^n b^{n -1}_{sj} - 3 a^{n -1}_{sj})$.}

We also need to verify that the arc capacities are non-negative: 
\begin{lemma}  \lemlab{phi}
	For $n \geq 2$ and $j < n$, the capacity $u^n_{nj}$ of the
	new internal arc $n \to j$ is non-negative.
\end{lemma}

\Proof:   
Figures \ref{fig:2cells} and \ref{fig:3cells} prove the statement for $n=2$ and $3$.  For $n \geq 4$, it suffices to show that $\newtheta^n b^{n-1}_{sj} - 3 a^{n-1}_{sj} \geq 0$, or equivalently $\newtheta^n \geq 3 a^{n-1}_{sj}/b_{sj}^{n-1}$ for $j <n$.

First, we claim that if $n \geq 4$, then $a^{n-1}_{sj}\ge  a^{n-1}_{s,j-1}$, i.e., the $a^{n-1}_{sj}$ are non-decreasing in $j$.  Since $a^{n-1}_{sj}= 4 a^{n-2}_{sj}$, this statement follows
by induction if we can establish that $a^{n-1}_{s,n-1} = \phi^{n-1}\ge
a^{n-1}_{s,n-2} = 4\phi^{n-2}$. By \LEMMA{phiGrowth}, $\phi^{n-2} \geq \phi^{2} = 4$ and moreover $\phi^{n-1} \geq 3( \phi^{n-2})^2 \geq 4 \phi^{n-2}$.

Next, we claim that $b^{n-1}_{sj}\le b^{n-1}_{s,j-1}$, i.e., the $b^{n-1}_{sj}$
are non-increasing in $j$.  Since $b^{n-1}_{s, n-1} = 1$ for all $n$ and
the $b^{n-1}_{sj} = (1+\newtheta^{n-1}) b^{n-2}_{sj}$, this follows by induction.

Using the above statements, \EQ{thdef}, and $b^{n-1}_{s,n-1} = 1$, we get:  
\begin{align*}
\max_{j < n} \; \frac{3a^{n-1}_{sj}}{b^{n-1}_{sj}} = \frac{3a^{n-1}_{s,n-1}}{
	b^{n-1}_{s,n-1}} = 3a^{n-1}_{s,n-1} = \newtheta^n. \qed
\end{align*}

\medskip

Finally, we prove that all $2^n$ possible cuts are min cuts on $\nscr^n$.  For $S \subseteq N^{n-1}$, we first consider cut $S$, with corresponding flow $x^{n,S}$ and $(\lam^{n,S}, \mu^{n,S}) \in (0, \frac{1}{4})\times
(\frac{\newtheta^n}{1+\newtheta^n}, 1)$.  
\begin{lemma}  \lemlab{xnon}
  Suppose $S \subseteq N^{n-1}$.  Flow $x^{n,S}$ is feasible at $(\lam^{n,S}, \mu^{n,S})$, and satisfies (SCS) with $S$, thus proving that $S$ is the unique min
  cut at $(\lam^{n,S}, \mu^{n,S})$ on $\nscr^n$.
\end{lemma}

\vcom{v17: Moved ``$x^{n,S}$ equals $x^{n-1,S}$ everywhere else'' from
  the first paragraph to the second paragraph, which deals with the
  internal arcs.  The first paragraph now focuses on the source/sink
  arcs, so I replaced the previous line with ``the saturation gaps of
  $s \to j$ and $j \to t$ do not change''; TM: further re-write to replace
  ``sat gaps'' with ``sat statuses'' on $s\to j$ and $j\to t$ }
\Proof: See Figures \ref{fig:nscr_minus} and \ref{fig:nscr_notInS}.
First consider (SCS) on source/sink arcs for
$1\le j < n$.  By \EQ{capacities}, \EQ{newdata_without},
\EQ{newflow_without}, and \EQ{Snon}, we increase $x^{n,S}_{sj},
u^{n,S}_{sj},x^{n,S}_{jt}$, and $u^{n,S}_{jt}$ by $\newtheta^n
b^{n-1}_{sj}$.  There is no flow on $n \to j$ and so conservation of
flow at $j$ is maintained.  Since the saturation statuses of $s \to j$
and $j \to t$ do not change, and $x^{n-1,S}$, $u^{n-1,S}$ satisfy (SCS), $x^{n,S}$ satisfies (SCS) on the source/sink arcs with $j< n$.  For $j = n$,
$\phi^{n}\ge 4$, $\lam^{n,S} < \frac{1}{4}$, and $\mu^{n,S} < 1$ imply
that:
\begin{align*}
u^n_{nt} = \frac{\phi^{n}}{2} \ge \frac{\phi^{n}}{4} + 1 > \phi^{n} \lam^{n,S} + \mu^{n,S} =
u^{n,S}_{sn} = x^{n,S}_{sn} = x^{n,S}_{nt}.
\end{align*} 
Hence $x^{n,S}$ is a feasible flow and the source/sink arcs satisfy (SCS).

Now consider (SCS) on internal arcs.  For $j <n$,
\EQ{capacities} and \EQ{newflow_without} imply that the original internal arcs have the same flows and capacities as $x^{n-1, S}, u^{n-1, S}$.  Hence, if $j\in S$, then all arcs
$j\to k$ ($k < j$) are saturated, which is necessary for (SCS) when
$k\notin S$.  Similarly, if $j\notin S$, then all arcs $j\to k$ have 0
flow, which is necessary for (SCS) when $k\in S$.  The new internal arcs $n \to j$ all have 0 flow.  Thus all conditions
in (SCS) are satisfied. \qed

Next, we consider cut $S+n$, with corresponding flow $x^{n, S+n}$ and  $(\lam^{n,S+n}, \mu^{n,S+n}) \in  (\frac{3}{4},1)\times (0, \frac{1}{1+\newtheta^n})$.  
\begin{lemma}  \lemlab{xyesn}
	Suppose $S \subseteq N^{n-1}$.  Flow $x^{n,S+n}$ is feasible at $(\lam^{n,S+n}, \mu^{n,S+n})$, and satisfies (SCS) with $S+n$, thus proving that $S+n$ is the unique min cut at $(\lam^{n,S+n}, \mu^{n,S+n})$ on $\nscr^n$.
\end{lemma}

\Proof: See Figures \ref{fig:nscr_minus} and \ref{fig:nscr_InS}.
Similar to Lemma \ref{lem:xnon}, first consider (SCS) on source/sink
arcs for $1 \leq j <n$.  By \EQ{capacities}, \EQ{newdata_with},
\EQ{newflow_with}, and \EQ{Syesn}, we add $3a^{n-1}_{sj}$ to
$x^{n-1,S}_{sj}$ and $u^{n-1,S}_{sj}$ to obtain $x^{n,S+n}_{sj}$ and
$u^{n,S+n}_{sj}$ , but we add $\theta^n b^{n-1}_{sj}$ to
$x^{n-1,S}_{jt}$ and $u^{n-1,S}_{jt}$ to obtain $x^{n,S+n}_{jt}$ and
$u^{n,S+n}_{jt}$.  To compensate for this update gap, the flow on $n \to j$ is $x^{n,S+n}_{nj }=\theta^n b^{n-1}_{sj} - 3a^{n-1}_{sj} = u^{n,S+n}_{nj}$; this ensures that $n \to j$ is saturated and conservation of flow at $j$ is maintained.  Again, the saturation statuses of $s \to j$ and $j \to t$ do not change, so the source/sink arcs of $j$ satisfy (SCS).  

For $j = n$, there is flow of $\phi^n/2 + \sum_{j<n} (\theta^n b_{sj}^{n-1} - 3a_{sj}^{n-1}) = \phi^n/2 + \sum_{j<n} u_{nj}^n$ going into $n$ on $s \to n$.  There is flow of $x_{nt}^{n,S+n} = \phi^n/2$ leaving $n$ to saturate $n \to t$, and flow of $x_{nj}^{n,S+n} = \theta^n b^{n-1}_{sj} - 3a^{n-1}_{sj}$ leaving $n$ to saturate $n \to j$.  Hence, conservation of flow holds at $n$, and it remains to check that $x_{sn}^{n,S+n}< u_{sn}^{n,S+n} = \phi^n \lambda^{n,S+n} + \mu^{n,S+n}$ to satisfy (SCS).  Since $\phi^{n} /4= \sum_{j < n}u^n_{nj}$ by \EQ{phidef}, and using $\lam^{n,S+n} > \frac{3}{4}$ and $\mu^{n,S+n} > 0$, we get the desired inequality as follows:
\begin{align*}
\phi^{n}\lam^{n,S+n}+ \mu^{n,S+n} > \phi^{n}\lam^{n,S+n} > \frac{3}{4}\phi^{n} 
= \phi^{n}/2 + \sum_{j < n}u^n_{nj}.
\end{align*}
Hence, $x^{n,S+n}$ is a feasible flow and all of the source/sink arcs satisfy (SCS).

The proof for the (SCS) conditions on internal arcs is similar to \LEMMA{xnon}, with the modification that the new internal arcs $n \to j$ are saturated.  \qed 

\medskip

\Proofof:{\relax \THEO{main}} Lemmas \ref{lem:xnon} and
\ref{lem:xyesn} show that for all $S\subseteq N^n$, $x^{n,S}$ and
$S$ satisfy (SCS) at $(\lam^{n,S}, \mu^{n,S})$, which proves that $S$ is
the unique min cut at $(\lam^{n,S}, \mu^{n,S})$.  \qed

\comment{This proves that
$\nscr^n$ has $2^n$ distinct min cuts, and so the complexity of SSM
max flow/min cut with two parameters is exponential.}

\section{Discussion} \seclab{discuss}

Parametric max flow/min cut networks with the SSM property have many
applications, and the GGT results and their extensions have proven to
be very fruitful.  It is disappointing that multi-parameter SSM max flow networks can have an exponential number of min cuts, even though they share the same nestedness property as single-parameter SSM max flow networks due to Topkis's result.

However, note that SSM and its attendant nestedness are
not the only way to get efficient parametric max flow algorithms.  For
example, the problem of ``max mean cut'' introduced in \cite{cmmc}
leads to a single parameter max flow problem that does not satisfy
nestedness (see \cite[Figure 3]{cmmc}).  Despite this, the Discrete
Newton Algorithm developed in
\cite{cmmc,radzik,radnewt,radchap,radchap2} is quite efficient, and
its framework can be extended to other non-max flow parametric
problems.  As another example, the extreme case of Chen's scheduling
problem \cite{chen} where each product has its own parameter can be
solved as a special case of the GGT framework, as shown by
\cite{ggtsch}.

Furthermore, solving multi-parameter problems does not necessarily
entail enumerating all of the cells.  Even if there are an exponential
number of cells, it might still be possible to get a polynomial-time
algorithm to solve some parametric problems.  One example is the network
interdiction problem with multiple budgets, where \cite{budgetnet}
derives a polynomial-time recursive algorithm that solves the problem
for any fixed number of parameters (budgets).

As a corollary of \LEMMA{phiGrowth}, we can derive fairly
tight bounds on the size of the data in our instances:

\begin{coro}  \corolab{dexp}
	For $n \geq 3$, 
\begin{align*}
\frac{1}{3}\cdot 2^{(2^{n})} \le \phi^{n}\le   \frac{1}{17}\cdot 2^{(2^{n+.5})} .
\end{align*}
\end{coro}
\Proof: We use induction on $n$.  For the lower bound, the base of the
induction is $\phi^{3} = 144 \geq \frac{1}{3}\cdot 2^{(2^{3})}$.  Using
\LEMMA{phiGrowth}, if $n \ge 4$, then
$\phi^{n}\ge 3\big(\phi^{n-1}\big)^2 \ge 3\big(\frac{1}{3}\cdot
2^{(2^{n-1})}\big)^2 = \frac{1}{3}\cdot 2^{(2^{n})}$, as desired.

\comment{
For the upper bound, the base of the induction is
$\phi^{3} = 144 \leq \frac{1}{6}\cdot 2^{(2^{3.3})}$.  When $n \ge4$, we can apply \EQ{key}.  In addition, \mbox{$\sum_{j < n-1} b^{n-2}_{sj} =  \left (\left(\phi^{n-1}/4\right) + 3\sum_{j<n-1} a^{n-2}_{sj}\right)/\newtheta^{n-1}$} by \LEMMA{phiGrowth}.  Along with the facts that $\newtheta^n = 3\phi^{n-1}$ and $\newtheta^{n-1} \geq 12$ for $n >3$, we obtain:
\begin{align}
\phi^{n} & =   3 (\phi^{n-1})^2 +  4\sum_{j<n-1} \big[ \newtheta^nb^{n-2}_{sj} +
3a^{n-2}_{sj}(\newtheta^n - 4)\big]  \nonumber \\
& \le    3 (\phi^{n-1})^2+ 4\newtheta^n \left( \sum_{j < n-1} b^{n-2}_{sj} + 3\sum_{j < n-1}a^{n-2}_{sj}\right)  \nonumber \\
& \leq 3 (\phi^{n-1})^2+ \phi^{n-1}\left( \frac{\phi^{n-1}}{4} +   39\sum_{j<n-1}a^{n-2}_{sj} \right )\eqlab{phiup}. 
\end{align}

As a sub-claim, we can show that $\sum_{j<n'} a^{n'-1}_{sj} \leq \phi^{n'}/18$ for $n' \geq 3$ by induction.  In the base case, $\phi^{3} = 144 = 18\sum_{j<3} a^{2}_{sj}$.  For $n'>3$, $\phi^{n'-1} \geq 144$ implies:  
\begin{align*}
18\sum_{j<n'} a^{n'-1}_{sj}& = 18 \left( 4 \sum_{j<n'-1} a^{n'-2}_{sj} + \phi^{n'-1} \right)
\\ & \leq 18\left( \frac{4}{18} \phi^{n'-1} + \phi^{n'-1} \right) 
\\ &  = 22 \phi^{n'-1} \leq 3 \left(\phi^{n'-1} \right)^2 \leq  \phi^{n'}.
\end{align*}

Plugging the sub-claim into \EQ{phiup} yields the desired bound:
\begin{align*}
\phi^{n} & \le 3 (\phi^{n-1})^2+ \phi^{n-1}\left( \frac{\phi^{n-1}}{4} +   \frac{39 \phi^{n-1}}{18} \right )\\
&<6\left(\phi^{n-1} \right)^2
\le 6\left(\frac{1}{6}2^{(2^{n-1-0.3})}\right)^2 
= \frac{1}{6} 2^{(2^{n+0.3})}. \qed
\end{align*}}

For the upper bound the base of the induction is
$\phi^{3} = 144\le \frac{1}{17}\cdot 2^{(2^{3+.5})} = 149.73$.  For $n \ge4$, we re-write
\EQ{key} \comment{using the recursive fact that
$a^{n-2}_{sj} = 4^{n-2-j}\phi^{j} $} as:
\begin{align}
        \phi^{n} & =   3(\phi^{n-1})^2 + 4\sum_{j < n-1}
                 \left[ \newtheta^n b^{n-2}_{sj} + 3(\newtheta^n
                 -4)a^{n-2}_{sj}\right]  \nonumber \\
        & \le   3(\phi^{n-1})^2 + 4\newtheta^n\sum_{j < n-1}
                 \left[  b^{n-2}_{sj} + 3a^{n-2}_{sj}\right]
          \nonumber \\
      & = 3\phi^{n-1}\bigg[ \phi^{n-1} + 4\big(\sum_{j<n-1}
        b^{n-2}_{sj} + \sum_{j<n-1} 3a^{n-2}_{sj}\big) \bigg]. \eqlab{phiup}
\end{align}
We separately derive upper bounds on $\sum_{j<n-1} 3a^{n-2}_{sj}$
and $\sum_{j<n-1} b^{n-2}_{sj}$ to finish the proof.

By induction,
$a^{n-2}_{sj} = 4^{n-2-j}\phi^{j}$, and so
$a^{n-2}_{s,j-1}/a^{n-2}_{sj} = 4\phi^{j-1}/\phi^{j}$.
If $j \geq 3$, then \LEMMA{phiGrowth} showed that $\phi^{j}\ge 3(\phi^{j-1})^2$, or
$4\phi^{j-1}/\phi^{j}\le 4/(3\phi^{j-1})$.  Since $\phi^{j-1} \geq \phi^{2} = 4$, we get
$4\phi^{j-1}/\phi^{j}\le \frac{1}{3}$ (or $\phi^{j-1} \leq \frac{1}{12}\phi^{j}$).  Since each term of
$\sum_{2\le j<n-1} a^{n-2}_{sj}$ is at most a third of the previous
term, the sum is less than $\frac{3}{2}$ times its leading term
(namely $a^{n-2}_{s,n-2} = \phi^{n-2}$), or
$ \sum_{2\le j<n-1} 3a^{n-2}_{sj}\le \frac{9}{2}\phi^{n-2} \le
\frac{3}{8}\phi^{n-1}$.  Finally,
$a^{n-2}_{s1} = 4^{n-3} =  4^{n-4}\phi^{2} \le  4^{n-4} \gamma^{n-1}/ 12^{n-3} \le
 \frac{1}{12}\phi^{n-1}$, and we get
\begin{equation}
  \sum_{1\le j<n-1} 3a^{n-2}_{sj}\le \phi^{n-1}.  \eqlab{suma}
\end{equation}

Now \EQ{phidef} and $\theta^{n-1} = 3 \phi^{n-2} \geq 12$
imply that:
\vcom{Replaced $a_{sj}^{n-1}$ in first line with $a_{sj}^{n-2}$.  I think the first line is correct now based on (4).}
\begin{align*}
  \sum_{j<n-1} b^{n-2}_{sj} & = \frac{1}{\newtheta^{n-1}}\left[
                              \frac{\phi^{n-1}}{4} + \sum_{j<n-1} 3
                              a^{n-2}_{sj}\right] \\
  & \le \frac{1}{12} \left[ \frac{\phi^{n-1}}{4} + \sum_{j<n-1}
                              3a^{n-2}_{sj} \right] \\
  & \le \frac{1}{12} \left[ \frac{\phi^{n-1}}{4} + 
  \phi^{n-1} \right]  = \frac{5}{48}\phi^{n-1}.
\end{align*}

Plugging this and \EQ{suma} into \EQ{phiup} and using induction yields
\begin{align*}
  \phi^{n} & \le 3\phi^{n-1}\left[ \phi^{n-1} + 4\bigg(\frac{5}{48}\phi^{n-1} +
           \phi^{n-1}\bigg)\right]  \\
  & < 17(\phi^{n-1})^2\le 17\bigg(\frac{1}{17}
  2^{(2^{(n-1)+.5})}\bigg)^2 = \frac{1}{17} 2^{(2^{n+.5})},
\end{align*}
as desired.  \qed


\CORO{dexp} shows that the coefficients $a^n_{sj}$, $b^n_{sj}$ arising in the construction of $\nscr^n$ have an exponential number of bits.
It is natural to wonder whether there is a more parsimonious
construction that still has an exponential number of min cuts, but
where all data have a polynomial number of bits.
An easy induction shows that network $\nscr^n$
has $(n^2 + n + 2)/2 = \Theta(n^2)$ arcs.  
Could one get a lesser bound on the number of cells from networks
which are less dense?
\vcom{Above: I replaced the $\theta(n^2)$ with $\Theta(n^2)$ ).  This
  was $\Theta$ in an early draft, but it was changed to $\theta$, so
  please verify! TM: correct}

Note that our construction is inherently asymmetric between $\lam$ and
$\mu$.  This comes from the construction of node $n$ and its arcs:  we require a large $a^n_{sn}$ and small $b^n_{sn}$ to ensure that node $n$ does not belong to the min cut when $\lam$ is small, and subsequently enters the min cut when $\lam$ is large.  This asymmetry propagates throughout the whole construction.  Is there a more aesthetic construction that is symmetric between $\lam$ and $\mu$?

It is worthwhile to point out that SSM still implies some nice
behavior of min cuts.  For example, if an algorithm follows
any ``northeast'' path in the $(\lam, \mu)$-plane (i.e., that never
reduces the $\lam$ coordinate or the $\mu$ coordinate), then Topkis's result  implies that the path will encounter at most $n$ min cuts.

When we were doing exploratory computations at the beginning of this
project, we generated many small instances.  Empirically it appears that
``most'' instances have only a small number of min cuts.  In our
construction that forces $2^n$ cells, most of the cells are vanishingly small and concentrated in a very small region (this
behavior is likely connected to the exponentially large data used in
our construction).  This behavior can be understood via {\em smoothed
  analysis}.  For example, results in \cite{smoothed} show that under a
slight perturbation of the coefficients, the expected
number of cells is polynomial in $n$.

\textbf{Funding:}  The research
of the second and third authors were partially supported by NSERC
Discovery Grants; the research of the first author was
supported by USRA Grants.

\textbf{Acknowledgements:}  We thank Joseph Cheriyan and
Richard Anstee for bringing us together; Joseph Paat for his
comments on an earlier draft; and an anonymous referee for their
careful reading, and many good suggested improvements, including
pointing out \cite{smoothed}.

\bibliographystyle{agsm}
\bibliography{bibParam}

\end{document}

%% file: Fig3_Nminus.pdf_t
\begin{picture}(0,0)%
\includegraphics{Fig3_Nminus.pdf}%
\end{picture}%
\setlength{\unitlength}{3947sp}%
\begingroup\makeatletter\ifx\SetFigFont\undefined%
\gdef\SetFigFont#1#2#3#4#5{%
  \reset@font\fontsize{#1}{#2pt}%
  \fontfamily{#3}\fontseries{#4}\fontshape{#5}%
  \selectfont}%
\fi\endgroup%
\begin{picture}(7802,3671)(-3941,-920)
\put(3601, 14){\makebox(0,0)[b]{\smash{{\SetFigFont{12}{14.4}{\rmdefault}{\mddefault}{\updefault}{\color[rgb]{0,0,0}$t$}%
}}}}
\put(841,914){\makebox(0,0)[lb]{\smash{{\SetFigFont{10}{12.0}{\rmdefault}{\mddefault}{\updefault}{\color[rgb]{0,0,0}$x^{-}=u^{-}$}%
}}}}
\put(841,-511){\makebox(0,0)[lb]{\smash{{\SetFigFont{10}{12.0}{\rmdefault}{\mddefault}{\updefault}{\color[rgb]{0,0,0}$x^{-}<u^{-}$}%
}}}}
\put(-2699,-511){\makebox(0,0)[lb]{\smash{{\SetFigFont{10}{12.0}{\rmdefault}{\mddefault}{\updefault}{\color[rgb]{0,0,0}$x^{-}=u^{-}$}%
}}}}
\put(685,1664){\makebox(0,0)[rb]{\smash{{\SetFigFont{10}{12.0}{\rmdefault}{\mddefault}{\updefault}{\color[rgb]{0,.56,0}\green{$x^-=u^-$}}%
}}}}
\put(  1,2264){\makebox(0,0)[b]{\smash{{\SetFigFont{12}{14.4}{\rmdefault}{\mddefault}{\updefault}{\color[rgb]{0,0,0}$k$}%
}}}}
\put(  1,764){\makebox(0,0)[b]{\smash{{\SetFigFont{12}{14.4}{\rmdefault}{\mddefault}{\updefault}{\color[rgb]{0,0,0}$j\in S$}%
}}}}
\put(-3599, 14){\makebox(0,0)[b]{\smash{{\SetFigFont{12}{14.4}{\rmdefault}{\mddefault}{\updefault}{\color[rgb]{0,0,0}$s$}%
}}}}
\put(-2699,914){\makebox(0,0)[lb]{\smash{{\SetFigFont{10}{12.0}{\rmdefault}{\mddefault}{\updefault}{\color[rgb]{0,0,0}$x^{-}<u^{-}$}%
}}}}
\put(-599,119){\makebox(0,0)[lb]{\smash{{\SetFigFont{10}{12.0}{\rmdefault}{\mddefault}{\updefault}{\color[rgb]{.82,0,0}$x^-=0$}%
}}}}
\put(  1,-736){\makebox(0,0)[b]{\smash{{\SetFigFont{12}{14.4}{\rmdefault}{\mddefault}{\updefault}{\color[rgb]{0,0,0}$i\notin S$}%
}}}}
\end{picture}%

%% file: Fig4_N_notInS.pdf_t
\begin{picture}(0,0)%
\includegraphics{Fig4_N_notInS.pdf}%
\end{picture}%
\setlength{\unitlength}{3947sp}%
\begingroup\makeatletter\ifx\SetFigFont\undefined%
\gdef\SetFigFont#1#2#3#4#5{%
  \reset@font\fontsize{#1}{#2pt}%
  \fontfamily{#3}\fontseries{#4}\fontshape{#5}%
  \selectfont}%
\fi\endgroup%
\begin{picture}(7847,5084)(-3986,-2420)
\put(781,914){\makebox(0,0)[lb]{\smash{{\SetFigFont{10}{12.0}{\rmdefault}{\mddefault}{\updefault}{\color[rgb]{0,0,0}add $\blue{\theta b^{-}}$ to both sides${}\Rightarrow x^+=u^+$}%
}}}}
\put(781,-286){\makebox(0,0)[lb]{\smash{{\SetFigFont{10}{12.0}{\rmdefault}{\mddefault}{\updefault}{\color[rgb]{0,0,0}$x^{-}<u^{-}$}%
}}}}
\put(-983, -1){\makebox(0,0)[lb]{\smash{{\SetFigFont{10}{12.0}{\rmdefault}{\mddefault}{\updefault}{\color[rgb]{.82,0,0}$x^+=x^-=0$}%
}}}}
\put(1081,-1936){\makebox(0,0)[lb]{\smash{{\SetFigFont{10}{12.0}{\rmdefault}{\mddefault}{\updefault}{\color[rgb]{.56,0,.56}$x^+<u^+=\phi/2$}%
}}}}
\put(781,-511){\makebox(0,0)[lb]{\smash{{\SetFigFont{10}{12.0}{\rmdefault}{\mddefault}{\updefault}{\color[rgb]{0,0,0}add $\blue{\theta b^{-}}$ to both sides${}\Rightarrow x^+<u^+$}%
}}}}
\put(-3119,914){\makebox(0,0)[lb]{\smash{{\SetFigFont{10}{12.0}{\rmdefault}{\mddefault}{\updefault}{\color[rgb]{0,0,0}add $\blue{\theta b^{-}}$ to both sides${}\Rightarrow x^+<u^+$}%
}}}}
\put(-2399,-1936){\makebox(0,0)[lb]{\smash{{\SetFigFont{10}{12.0}{\rmdefault}{\mddefault}{\updefault}{\color[rgb]{.56,0,.56}$x^+ = u^+ \approx \phi/4$}%
}}}}
\put(-3119,-511){\makebox(0,0)[lb]{\smash{{\SetFigFont{10}{12.0}{\rmdefault}{\mddefault}{\updefault}{\color[rgb]{0,0,0}add $\blue{\theta b^{-}}$ to both sides${}\Rightarrow x^+=u^+$}%
}}}}
\put(1561,1664){\makebox(0,0)[rb]{\smash{{\SetFigFont{10}{12.0}{\rmdefault}{\mddefault}{\updefault}{\color[rgb]{0,.56,0}\green{$x^+=x^-=u^-=u^+$}}%
}}}}
\put(601,-1381){\makebox(0,0)[rb]{\smash{{\SetFigFont{10}{12.0}{\rmdefault}{\mddefault}{\updefault}{\color[rgb]{.82,0,0}$x^+=0$}%
}}}}
\put(  1,2264){\makebox(0,0)[b]{\smash{{\SetFigFont{12}{14.4}{\rmdefault}{\mddefault}{\updefault}{\color[rgb]{0,0,0}$k$}%
}}}}
\put(-3119,-286){\makebox(0,0)[lb]{\smash{{\SetFigFont{10}{12.0}{\rmdefault}{\mddefault}{\updefault}{\color[rgb]{0,0,0}$x^{-}=u^{-}$}%
}}}}
\put(-3119,1139){\makebox(0,0)[lb]{\smash{{\SetFigFont{10}{12.0}{\rmdefault}{\mddefault}{\updefault}{\color[rgb]{0,0,0}$x^{-}<u^{-}$}%
}}}}
\put(781,1139){\makebox(0,0)[lb]{\smash{{\SetFigFont{10}{12.0}{\rmdefault}{\mddefault}{\updefault}{\color[rgb]{0,0,0}$x^{-}=u^{-}$}%
}}}}
\put(-3599, 14){\makebox(0,0)[b]{\smash{{\SetFigFont{12}{14.4}{\rmdefault}{\mddefault}{\updefault}{\color[rgb]{0,0,0}$s$}%
}}}}
\put(3601, 14){\makebox(0,0)[b]{\smash{{\SetFigFont{12}{14.4}{\rmdefault}{\mddefault}{\updefault}{\color[rgb]{0,0,0}$t$}%
}}}}
\put(  1,-2236){\makebox(0,0)[b]{\smash{{\SetFigFont{12}{14.4}{\rmdefault}{\mddefault}{\updefault}{\color[rgb]{0,0,0}$n\notin S$}%
}}}}
\put(  1,764){\makebox(0,0)[b]{\smash{{\SetFigFont{12}{14.4}{\rmdefault}{\mddefault}{\updefault}{\color[rgb]{0,0,0}$j\in S$}%
}}}}
\put(  1,-736){\makebox(0,0)[b]{\smash{{\SetFigFont{12}{14.4}{\rmdefault}{\mddefault}{\updefault}{\color[rgb]{0,0,0}$i\notin S$}%
}}}}
\end{picture}%

%% file: Fig5_N_inS.pdf_t
\begin{picture}(0,0)%
\includegraphics{Fig5_N_inS.pdf}%
\end{picture}%
\setlength{\unitlength}{3947sp}%
\begingroup\makeatletter\ifx\SetFigFont\undefined%
\gdef\SetFigFont#1#2#3#4#5{%
  \reset@font\fontsize{#1}{#2pt}%
  \fontfamily{#3}\fontseries{#4}\fontshape{#5}%
  \selectfont}%
\fi\endgroup%
\begin{picture}(7817,5151)(-3956,-2487)
\put(1081,-1936){\makebox(0,0)[lb]{\smash{{\SetFigFont{10}{12.0}{\rmdefault}{\mddefault}{\updefault}{\color[rgb]{.56,0,.56}$x^+=u^+=\phi/2$}%
}}}}
\put(-3119,1139){\makebox(0,0)[lb]{\smash{{\SetFigFont{10}{12.0}{\rmdefault}{\mddefault}{\updefault}{\color[rgb]{0,0,0}$x^{-}<u^{-}$}%
}}}}
\put(781,1139){\makebox(0,0)[lb]{\smash{{\SetFigFont{10}{12.0}{\rmdefault}{\mddefault}{\updefault}{\color[rgb]{0,0,0}$x^{-}=u^{-}$}%
}}}}
\put(781,-286){\makebox(0,0)[lb]{\smash{{\SetFigFont{10}{12.0}{\rmdefault}{\mddefault}{\updefault}{\color[rgb]{0,0,0}$x^{-}<u^{-}$}%
}}}}
\put(781,914){\makebox(0,0)[lb]{\smash{{\SetFigFont{10}{12.0}{\rmdefault}{\mddefault}{\updefault}{\color[rgb]{0,0,0}add $\blue{\theta b^{-}}$ to both sides${}\Rightarrow x^+=u^+$}%
}}}}
\put(-3119,-511){\makebox(0,0)[lb]{\smash{{\SetFigFont{10}{12.0}{\rmdefault}{\mddefault}{\updefault}{\color[rgb]{0,0,0}add $\blue{3a^{-}}$ to both sides${}\Rightarrow x^+=u^+$}%
}}}}
\put(-983, -1){\makebox(0,0)[lb]{\smash{{\SetFigFont{10}{12.0}{\rmdefault}{\mddefault}{\updefault}{\color[rgb]{.82,0,0}$x^+=x^-=0$}%
}}}}
\put(-3119,914){\makebox(0,0)[lb]{\smash{{\SetFigFont{10}{12.0}{\rmdefault}{\mddefault}{\updefault}{\color[rgb]{0,0,0}add $\blue{3a^{-}}$ to both sides${}\Rightarrow x^+<u^+$}%
}}}}
\put(781,-511){\makebox(0,0)[lb]{\smash{{\SetFigFont{10}{12.0}{\rmdefault}{\mddefault}{\updefault}{\color[rgb]{0,0,0}add $\blue{\theta b^{-}}$ to both sides${}\Rightarrow x^+<u^+$}%
}}}}
\put(1561,1664){\makebox(0,0)[rb]{\smash{{\SetFigFont{10}{12.0}{\rmdefault}{\mddefault}{\updefault}{\color[rgb]{0,.56,0}\green{$x^+=x^-=u^-=u^+$}}%
}}}}
\put(1681,-1381){\makebox(0,0)[rb]{\smash{{\SetFigFont{10}{12.0}{\rmdefault}{\mddefault}{\updefault}$\green{x^+=u^+=}\blue{\theta b^- -3a^-}$}}}}
\put(-2399,-1936){\makebox(0,0)[lb]{\smash{{\SetFigFont{10}{12.0}{\rmdefault}{\mddefault}{\updefault}{\color[rgb]{.56,0,.56}$x^+<u^+ \approx 3\phi/4$}%
}}}}
\put(  1,2264){\makebox(0,0)[b]{\smash{{\SetFigFont{12}{14.4}{\rmdefault}{\mddefault}{\updefault}{\color[rgb]{0,0,0}$k$}%
}}}}
\put(-3119,-286){\makebox(0,0)[lb]{\smash{{\SetFigFont{10}{12.0}{\rmdefault}{\mddefault}{\updefault}{\color[rgb]{0,0,0}$x^{-}=u^{-}$}%
}}}}
\put(  1,-736){\makebox(0,0)[b]{\smash{{\SetFigFont{12}{14.4}{\rmdefault}{\mddefault}{\updefault}{\color[rgb]{0,0,0}$i\notin S$}%
}}}}
\put(  1,764){\makebox(0,0)[b]{\smash{{\SetFigFont{12}{14.4}{\rmdefault}{\mddefault}{\updefault}{\color[rgb]{0,0,0}$j\in S$}%
}}}}
\put(  1,-2236){\makebox(0,0)[b]{\smash{{\SetFigFont{12}{14.4}{\rmdefault}{\mddefault}{\updefault}{\color[rgb]{0,0,0}$n\in S$}%
}}}}
\put(-3599, 14){\makebox(0,0)[b]{\smash{{\SetFigFont{12}{14.4}{\rmdefault}{\mddefault}{\updefault}{\color[rgb]{0,0,0}$s$}%
}}}}
\put(3601, 14){\makebox(0,0)[b]{\smash{{\SetFigFont{12}{14.4}{\rmdefault}{\mddefault}{\updefault}{\color[rgb]{0,0,0}$t$}%
}}}}
\end{picture}%